# The observation of in-plane quantum Griffiths singularity in two-dimensional crystalline superconductors

Yi Liu[1,2], Shichao Qi[1], Jingchao Fang[1], Jian Sun[1], Chong Liu[3], Yanzhao Liu[1], Junjie Qi[4], Ying Xing[5], Haiwen Liu[6,†], Xi Lin[1,4,7], Lili Wang[3], Qi-Kun Xue[3,4], X. C. Xie[1,4,7], Jian Wang[1,4,7,*]

[1]*International Center for Quantum Materials, School of Physics, Peking University, Beijing 100871, China.*
[2]*Department of Physics, Renmin University of China, Beijing 100872, China.*
[3]*State Key Laboratory of Low-Dimensional Quantum Physics, Department of Physics, Tsinghua University, Beijing 100084, China.*
[4]*Beijing Academy of Quantum Information Sciences, Beijing 100193, China.*
[5]*Department of Materials Science and Engineering, College of New Energy and Materials, China University of Petroleum, Beijing 102249, China.*
[6]*Center for Advanced Quantum Studies, Department of Physics, Beijing Normal University, Beijing 100875, China.*
[7]*CAS Center for Excellence in Topological Quantum Computation, University of Chinese Academy of Sciences, Beijing 100190, China.*

Quantum Griffiths singularity (QGS) reveals the profound influence of quenched disorder on the quantum phase transitions, characterized by the divergence of the dynamical critical exponent at the boundary of the vortex glass-like phase, named as quantum Griffiths phase. However, in the absence of vortices, whether the QGS can exist under parallel magnetic field remains a puzzle. Here we study the magnetic field induced superconductor-metal transition in ultrathin crystalline $PdTe_2$ films grown by molecular beam epitaxy. Remarkably, the QGS emerges under both perpendicular and parallel magnetic field in 4-monolayer $PdTe_2$ films. The direct activated scaling analysis with a new irrelevant correction has been proposed, providing important evidence of QGS. With increasing film thickness to 6 monolayers, the QGS disappears under perpendicular field but persists under parallel field, and this discordance may originate from the differences in microscopic processes. Our work demonstrates the universality of parallel field induced QGS and can stimulate further investigations on novel quantum phase transitions under parallel magnetic field.

Two-dimensional (2D) crystalline superconductors [1] are ideal platforms to study the quantum phase transition, a continuous phase transition at absolute zero temperature [2,3]. As a prototype of quantum phase transition, the superconductor-insulator/metal transition (SIT/SMT) has been widely and intensely investigated, where the quantum fluctuations play a dominant role and determine its characteristics [4,5]. The recent observations of quantum Griffiths singularity (QGS) of SMT in low-dimensional superconducting systems [6-13], characterized by a divergent critical exponent $zv$, challenge the conventional understanding of the quantum phase transition. QGS reveals the profound influence of the quenched disorder on SMT, which originates from the disorder driven evolution from vortex lattice to vortex glass-like phase, named as quantum Griffiths phase. The quantum Griffiths phase consists of large superconducting rare regions and the surrounding normal state. The size of these superconducting regions keeps increasing with decreasing temperature to zero, and the slow dynamics leads to a divergent critical exponent $zv$ of SMT in 2D systems [14,15], in contrast to the constant $zv$ observed in conventional quantum phase transitions [2,3]. The previous experimental works focus on the observation of QGS under perpendicular magnetic field. Under parallel field, the experimental investigation of QGS in the absence of vortices in 2D crystalline superconductors is highly desired.

In this paper, we report the transport properties of 4-ML and 6-ML $PdTe_2$ films via ultralow temperature transport measurement. Remarkably, the divergence of critical exponent $zv$ as an evidence of QGS is detected in 4-ML $PdTe_2$ films under both perpendicular and parallel magnetic field. Moreover, the QGS is directly identified by the activated scaling analysis with a new irrelevant correction. Interestingly, with increasing film thickness, the QGS disappears in the 6-ML film under perpendicular field but still exists under parallel field, revealing different microscopic processes of QGS under different field directions. We propose that the disorder can significantly influence the strength of spin-orbit coupling (SOC) and the in-plane critical field, which gives rise to the quantum Griffiths phase without vortex formation.

The ultrathin crystalline PdTe$_2$ films were epitaxially grown on Nb-doped SrTiO$_3$(001) substrates in the ultrahigh vacuum molecular beam epitaxy chamber (See Methods for details [16]). The morphology of the PdTe$_2$ films is characterized by the scanning tunneling microscope [20]. The PdTe$_2$ thin films are ambient-stable superconductors, which do not require capping layer for *ex situ* transport measurement. Figure 1 presents the superconducting properties of 4-ML PdTe$_2$ film, measured in a dilution refrigerator (MNK 126-450; Leiden Cryogenics BV) down to 20 mK. The standard four-electrode transport measurements are schematically shown in the inset of Fig. 1(c) (See Methods for details [16]). The superconducting transition begins at $T_c^{\text{onset}} = 700$ mK, which is defined as the crossing point of the linear extrapolations of normal state and superconducting transition curve. With decreasing temperature, the sheet resistance drops to zero within the measurement resolution at $T_c^{\text{zero}} = 570$ mK. As the perpendicular magnetic field increases, the 4-ML PdTe$_2$ film undergoes a superconductor to weakly localized metal transition with a quantum critical resistance (around 977 Ω) much smaller than the quantum resistance for Cooper pairs ($h/4e^2 \sim 6.45$ $k\Omega$, where $h$ is the Planck constant and $e$ is the elementary charge), as shown in the inset of Fig. 1(a). The sheet resistance increases with decreasing temperature when the magnetic field exceeds 0.98 T, indicating localized metal behavior. The perpendicular magnetic field dependence of sheet resistance at different temperatures from 20 mK to 450 mK is displayed in Fig. 1(b) (See Fig. S1 for the magnetoresistance in a large magnetic field region). Different from conventional SMT, the magnetoresistance isotherms cross each other in a relatively large and well-defined transition region around 0.9 T at low temperatures rather than a single critical point, reminiscent of QGS. The crossing points of $R_s(B)$ curves at neighboring temperatures are shown as black dots in the inset of Fig. 1(b). Furthermore, based on the finite size scaling analysis [2,3,16], the magnetic field dependence of the effective "critical" exponent $zv$ is summarized in Fig. 1(c). When approaching the characteristic magnetic field $B_c^*$ and zero temperature, $zv$ grows rapidly and then diverges. The field dependence of $zv$ can be well fitted by the activated scaling law $zv \propto |B_c^* - B|^{-v\psi}$ with the correlation length exponent $v \approx 1.2$ and the tunneling critical exponent $\psi \approx 0.5$ for 2D systems [21,22], providing experimental evidence of QGS in the 4-ML PdTe$_2$ film under perpendicular magnetic field with the infinite-randomness quantum critical point. We define the QGS under perpendicular (out-of-plane) field as the out-of-plane QGS. The out-of-plane QGS is confirmed in another 4-ML PdTe$_2$ film as shown in Fig. S2.

The divergence of effective critical exponents $zv$ near the quantum critical point originates from the activated scaling behavior of QGS [23]. Thus, we utilize the direct activated scaling analysis with the irrelevant parameter correction as follows [13]: $R = \Phi\left(\left(\frac{B-B_c^*}{B_c^*}\right) \cdot \left(ln\frac{T^*}{T}\right)^{\frac{1}{v\psi}}, u \cdot \left(ln\frac{T^*}{T}\right)^{-y}\right)$. Here $v$ and $\psi$ are critical exponents, $T^*$ is the characteristic temperature of quantum fluctuation, $u$ is the leading irrelevant scaling variable and $y > 0$ is the associate irrelevant exponent. (See Supplemental Material Part II for detailed numerical scaling procedure) [16]. The irrelevant correction also gives a good fitting for the phase boundary $B_c(T)$ of superconductor-metal transition [13]: $\frac{B_c^* - B_c(T)}{B_c^*} \propto u \cdot \left(ln\frac{T^*}{T}\right)^{-\frac{1}{v\psi} - y}$. The fitting of $B_c(T)$ is shown in the inset of Fig. 1(b). The activated scaling of twenty-three sets of data in the temperature range from 20 mK to 240 mK are presented in Fig. 1(d), providing direct evidence of QGS.

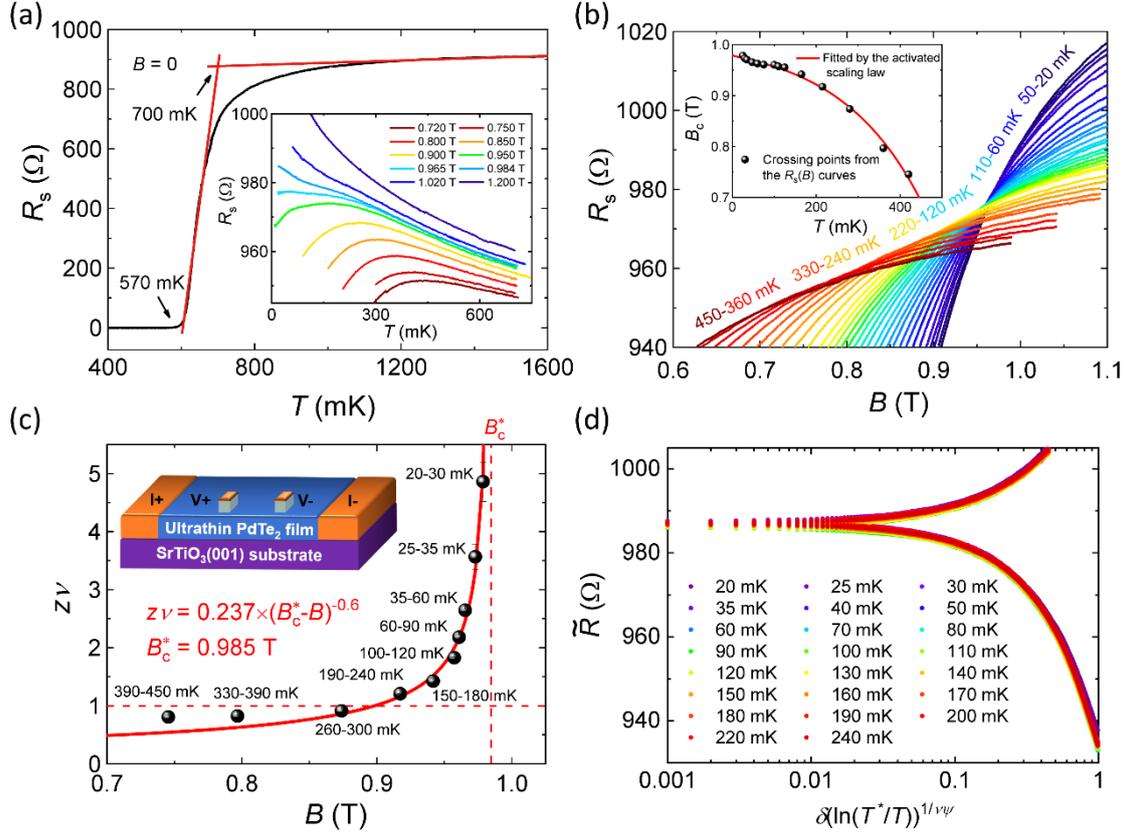

FIG. 1. The QGS of 4-ML PdTe$_2$ film under perpendicular magnetic field. (a) Temperature dependence of sheet resistance $R_s$ at zero magnetic field, showing $T_c^{onset} = 700$ mK and $T_c^{zero} = 570$ mK. Inset: $R_s(T)$ curves at various magnetic fields from 0.720 T to 1.200 T. (b) $R_s(B)$ curves at detailed temperatures ranging from 20 mK to 450 mK. Crossing points from the $R_s(B)$ curves are shown in the inset. The solid red line is the fitting curve from the activated scaling analysis with irrelevant correction. (c) Critical exponent $zv$ as a function of perpendicular field. The solid red line shows a fitting curve based on the activated scaling law and gives $B_c^* = 0.985$ T (vertical dashed line). The horizontal dashed red line shows $zv = 1$. Inset: The schematic for standard four-electrode transport measurements. (d) The direct activated scaling analysis of the $R_s(B)$ curves from 20 mK to 240 mK with the irrelevant correction. Here $\tilde{R}$ represents the sheet resistance considering the irrelevant correction.

We then investigate the superconducting properties of 4-ML PdTe$_2$ films under parallel magnetic fields up to 16 T in a commercial Physical Property Measurement System with dilution refrigerator option down to 50 mK. Interestingly, as shown in Fig. 2, the film exhibits the characteristics of QGS, which is quite similar to the observations under perpendicular field. To be specific, the $R_s(B)$ curves at different temperatures reveal a large transition region in Fig. 2(a) and the crossing points are consistent with the activated scaling model with irrelevant corrections as shown in Fig. 2(b). Moreover, the effective critical exponent $zv$ follows the activated scaling law $zv \propto |B_c^* - B|^{-0.6}$ when approaching characteristic field $B_c^*$ and zero temperature (Fig. 2(c)). The direct activated scaling analysis with irrelevant corrections (Fig. 2(d)) and the divergence of $zv$ provide solid evidence of QGS under parallel magnetic field (named as in-plane QGS).

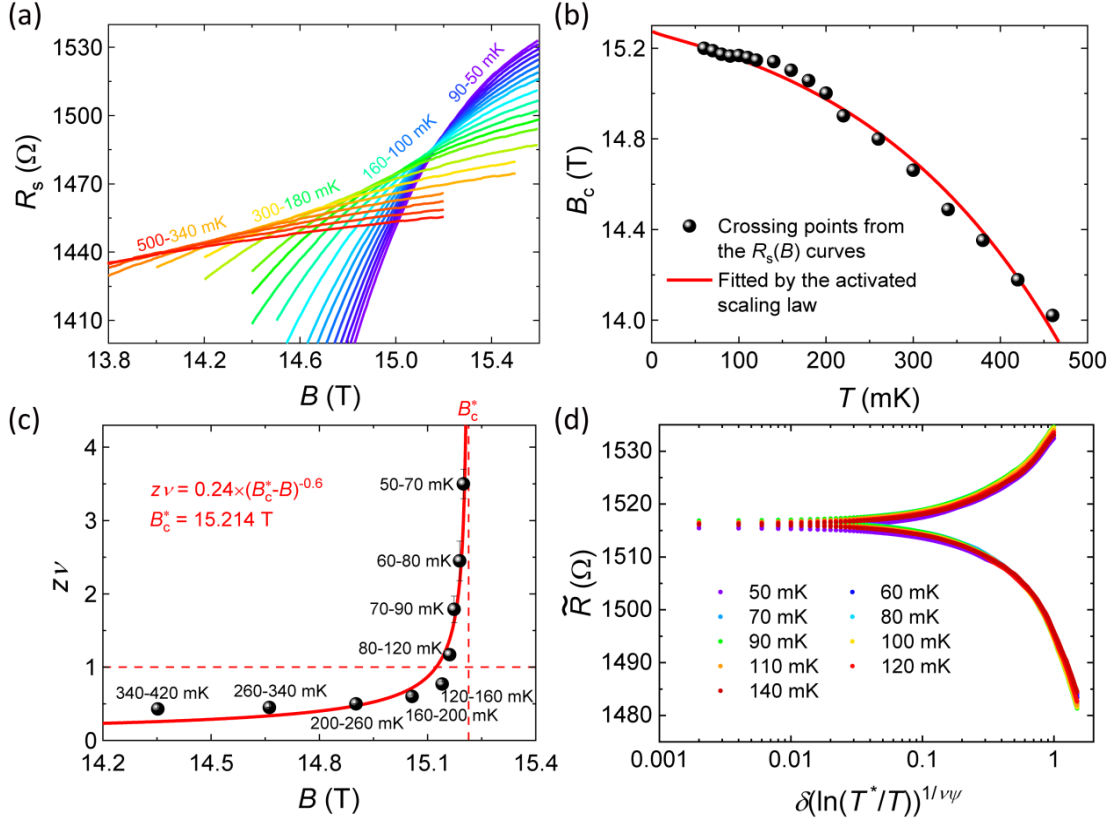

FIG. 2. The QGS of 4-ML PdTe$_2$ film under parallel magnetic field. (a) Parallel magnetic field dependence of $R_s$ at different temperatures. (b) Crossing points from the magnetoresistance isotherms. The solid red line is the fitting curve from the activated scaling analysis with irrelevant correction. (c) Critical exponent $zv$ obtained from finite size scaling analysis. The solid red line shows a fitting curve based on the activated scaling law and gives $B_c^* = 15.214$ T (vertical dashed line). The horizontal dashed red line shows $zv = 1$. (d) The direct activated scaling analysis of the $R_s(B)$ curves from 50 mK to 140 mK with the irrelevant correction. Here $\tilde{R}$ represents the sheet resistance considering the irrelevant correction.

The detection of the out-of-plane and in-plane QGS in 4-ML PdTe$_2$ films indicates the universality of QGS under different field orientations. It is noteworthy that the vortex-lattice phase can evolve into vortex-glass phase driven by quenched disorder under perpendicular magnetic field, which finally leads to QGS. However, this theoretical scenario does not work under parallel field where the vortex is absent, suggesting a different microscopic mechanism for in-plane QGS. Further exploration of QGS with different sample thickness may provide essential information for understanding the origin of in-plane QGS.

Thus, we performed ultralow temperature transport measurements on 6-ML PdTe$_2$ films. Figure 3(a) and 3(b) show the SMT behavior of the 6-ML PdTe$_2$ film under perpendicular field. During the SMT, the $R_s(T)$ curve at 1.8 T exhibits a plateau in a relatively large temperature regime at ultralow temperatures, corresponding to a single crossing point of the $R_s(B)$ curves below 220 mK. Finite size scaling in the inset of Fig. 3(b) further demonstrates a single value of $zv$ around 2.65 (the absolute value of the slope of the solid red line) between 90-220 mK (see details in Fig. S6), consistent with the quantum percolation theory [24-26]. However, $zv$ increases at lower temperatures below 90 mK. The above observation indicates that the characteristics of the out-of-plane QGS disappear with increasing film thickness, which very likely results from the relatively weak quantum fluctuation and disorder in thicker films. The Ioffe-Regel parameters are presented in Table S2 [16], revealing relatively weak disorder in the 6-ML PdTe$_2$ film. Interestingly, the main characteristic of in-plane QGS (i.e. the activate scaling law of $zv$) still persists in the same 6-ML PdTe$_2$ film under parallel magnetic field (Fig. 3(c) and 3(d)). The emergence of QGS is also confirmed by the direct activated scaling analysis with the irrelevant correction, as shown in the inset of Fig. 3(c) and Fig. S7.

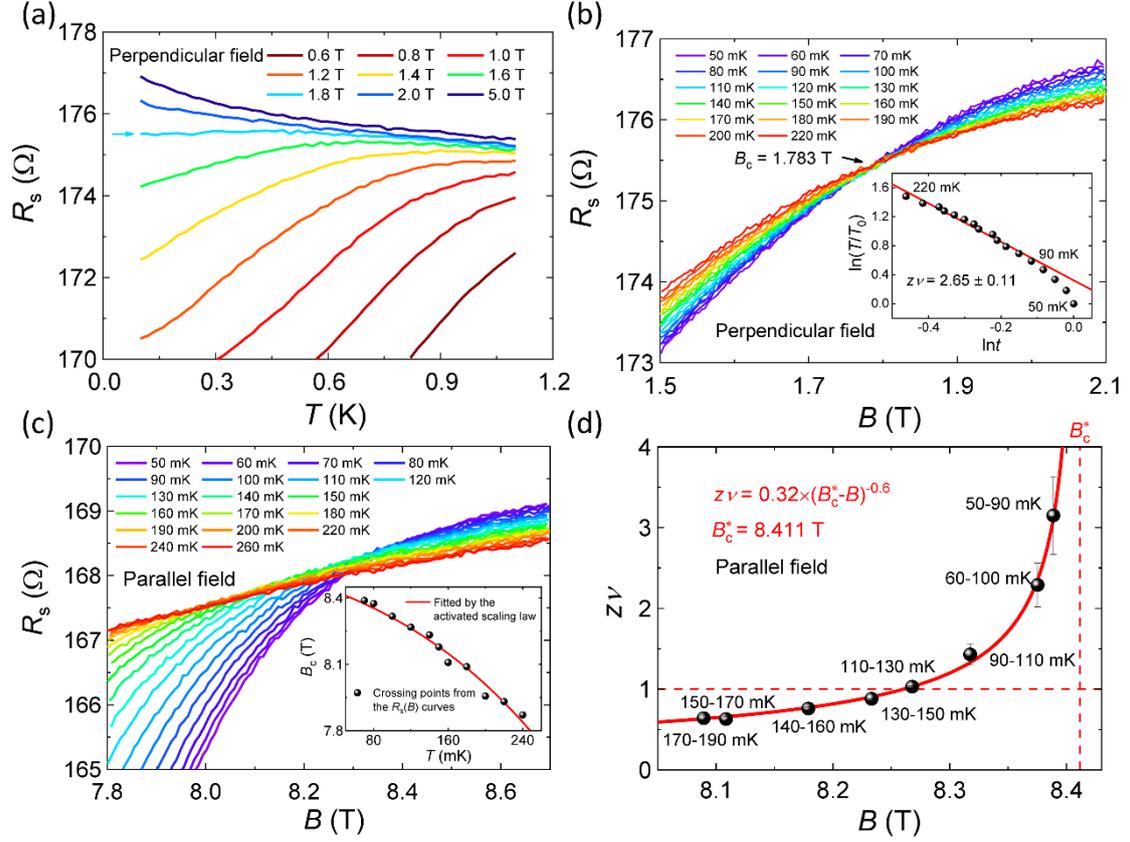

FIG. 3. Transport properties of 6-ML PdTe$_2$ film under perpendicular and parallel magnetic field. (a) $R_s(T)$ curves under perpendicular field ranging from 0.6 T to 5.0 T. The arrow indicates a plateau where $R_s$ remains nearly a constant at low temperatures. (b) $R_s(B)$ curves under perpendicular field at various temperatures, exhibiting one crossing point at $B_c = 1.783$ T. Inset: the temperature dependence of the scaling parameter $t$ ($t = (T/T_0)^{-1/zv}$). The solid red line is the linear fitting from 90 mK to 220 mK, showing $zv$ around 2.65. (c) The parallel magnetic field dependence of $R_s$ at different temperatures. Crossing points from $R_s(B)$ curves are shown in the inset. The solid red line is the fitting curve from the activated scaling analysis with irrelevant correction. (d) Critical exponent $zv$ obtained from scaling analysis. The solid red line shows a theoretical fitting based on the activated scaling law and gives $B_c^* = 8.411$ T (vertical dashed line). The horizontal dashed red line shows $zv = 1$.

This discordance may originate from the different microscopic processes driven by disorder under different magnetic field orientations. Under perpendicular magnetic field, the disorder effect can deform the vortex lattice and give rise to a vortex glass-like phase, where the large superconducting rare regions lead to the divergence of critical exponent $zv$. Compared to the QGS under perpendicular field, the different thickness dependent behavior as well as the absence of vortices in ultrathin 2D systems indicates a new mechanism of in-plane QGS. The mechanism of QGS without vortex has been theoretically revealed in superconducting nanowires, which can be extended to 2D superconductors [27]. Moreover, the ultrathin crystalline PdTe$_2$ films are type-II Ising superconductors with strong SOC [20,28]. The three-fold rotational symmetry of PdTe$_2$ films makes the effective field of Zeeman-type SOC along the out-of-plane direction, which protects the superconductivity under large in-plane magnetic field. The in-plane critical field of PdTe$_2$ films depends on the effective Zeeman-type SOC $\widetilde{\beta_{SO}} = \beta_{SO}/[1 + \hbar/(2\pi k_B T_c \tau_0)]$ [20,29]. Here $\beta_{SO}$ is the strength of Zeeman-type SOC, $\tau_0$ is the mean free time for spin-independent scattering and $T_c$ is the superconducting critical temperature. Because of the different local disorder strength with different value of $\tau_0$, the local effective SOC $\widetilde{\beta_{SO}}$ varies with location. When the in-plane magnetic field is near the mean-field critical field, the regions with relatively large disorder are easier to lose superconductivity and form the normal state, while the others keep superconducting and form the rare regions. The formation of rare regions under parallel field may give rise to the in-plane QGS. The disorder on the film surface and the interface between the film and the substrate may play a more important role in thinner films. Besides, the strain due to lattice mismatch between the film and substrate may lead to lattice distortion and finally contribute to the strength of disorder [30]. With decreasing film thickness, the strain is enhanced and thus the disorder strength is increased.

In conclusion, we systematically investigate the SMT behavior of ultrathin crystalline $PdTe_2$ films. Intriguingly, the QGS is observed in 4-ML $PdTe_2$ film under both perpendicular and parallel magnetic field. The evidence of QGS is also provided by a direct activated scaling analysis. With increasing film thickness, the out-of-plane QGS disappears while the in-plane QGS still exists in the 6-ML $PdTe_2$ film, indicating a new microscopic mechanism for the in-plane QGS. Our findings shed new light on the formation of QGS and inspire further investigations on the quantum phase transition under parallel magnetic field.

We thank Yaochen Li, Pengjie Wang, Cheng Chen, Zihan Cui for the help in transport measurement and data analysis. This work was financially supported by the National Key Research and Development Program of China (Grant No. 2018YFA0305604, No. 2017YFA0303300, No. 2017YFA0304600), the National Natural Science Foundation of China (Grant No. 11888101, No. 11774008, No. 11974430, No. 12022407), Beijing Natural Science Foundation (Z180010), the Strategic Priority Research Program of Chinese Academy of Sciences (Grant No. XDB28000000) and China Postdoctoral Science Foundation (Grant No. 2019M650290 and No. 2020T130021).

Yi L. and S.Q. contributed equally to this work.

*Corresponding author.

Jian Wang (jianwangphysics@pku.edu.cn)

†Corresponding author.

Haiwen Liu (haiwen.liu@bnu.edu.cn)

Supplemental Material for

# The observation of in-plane quantum Griffiths singularity in two-dimensional crystalline superconductors

Yi Liu[1,2], Shichao Qi[1], Jingchao Fang[1], Jian Sun[1], Chong Liu[3], Yanzhao Liu[1], Junjie Qi[4], Ying Xing[5], Haiwen Liu[6,†], Xi Lin[1,4,7], Lili Wang[3], Qi-Kun Xue[3,4], X. C. Xie[1,4,7], Jian Wang[1,4,7,*]

[1]*International Center for Quantum Materials, School of Physics, Peking University, Beijing 100871, China.*
[2]*Department of Physics, Renmin University of China, Beijing 100872, China.*
[3]*State Key Laboratory of Low-Dimensional Quantum Physics, Department of Physics, Tsinghua University, Beijing 100084, China.*
[4]*Beijing Academy of Quantum Information Sciences, Beijing 100193, China.*
[5]*Department of Materials Science and Engineering, College of New Energy and Materials, China University of Petroleum, Beijing 102249, China.*
[6]*Center for Advanced Quantum Studies, Department of Physics, Beijing Normal University, Beijing 100875, China.*
[7]*CAS Center for Excellence in Topological Quantum Computation, University of Chinese Academy of Sciences, Beijing 100190, China.*

Yi L. and S.Q. contributed equally to this work.
*Corresponding author.
Jian Wang (jianwangphysics@pku.edu.cn)
†Corresponding author.
Haiwen Liu (haiwen.liu@bnu.edu.cn)

## Contents

I. Methods
II. Discussions on quantum Griffiths singularity
III. Figures and Tables

## I. Methods

**Sample synthesis.** The well-prepared SrTiO$_3$(001) substrate is doped by 0.05 *wt*.% Nb. It was heated to 1100 °C to obtain the TiO$_2$-terminated surface in the ultrahigh vacuum. To grow high-quality PdTe$_2$ thin films, we evaporated Pd (99.995%) and Te (99.9999%) from Knudsen cells on the substrates at 120 °C. The flux ratio of Pd and Te is 1:2 and the growth rate are controlled at around 0.07 monolayer per minute. The morphology of the PdTe$_2$ film is characterized by the scanning tunneling microscopy at 4.8 K at a constant current mode and the bias voltage ($V_s$) is applied on the PdTe$_2$ film.

**Ultralow temperature transport measurements.** For *ex situ* transport measurements, we used standard four-electrode method to characterize the transport properties of the PdTe$_2$ films. Two indium strips ($I+$ and $I-$) were pressed along the width of the sample to make sure that the current could homogeneously pass through the film. The other two electrodes made by indium acted as the voltage probes ($V+$ and $V-$). The Hall resistance is measured by six-electrode method, as shown in Fig. S8. $V_H+$ and $V_H-$ are Hall electrodes. Besides, in our measurements under parallel magnetic field, the field is orthogonal to the excitation current. The resistance and magnetoresistance were measured in a dilution refrigerator (MNK 126-450; Leiden Cryogenics BV) and a commercial Physical Property Measurement System (Quantum Design, PPMS-16) with the dilution refrigerator option. The standard lock-in techniques were used in the measurements in MNK 126-450 system. The experimental data in Fig. 1 and S2 are obtained from two different 4-ML PdTe$_2$ films. The normal state resistance of these two films is around 500-900 Ω. The results in Fig. 2 and Fig. 1 are from the same film. However, the data in Fig. 2 were measured two years later than those in Fig. 1. Thus, the normal state resistance increases by 50.8%.

## II. Discussions on quantum Griffiths singularity.

Firstly, we provide finite size scaling analysis to obtain the effective critical exponents $zv$, and investigate the trend of $zv$ when approaching the zero temperature critical point. According to previous theoretical works [1,2], the critical sheet resistance of samples $R_c$ and magnetic field $B_c$ satisfies the scaling law expressed as: $R_s(B,T) = R_c \cdot F(|B - B_c|T^{-1/zv})$, where $F$ is an arbitrary function with $F(0) = 1$, $z$ and $v$ are the dynamical critical exponent and correlation length exponent, respectively. As mentioned in the main text, the crossing points at adjacent isotherms change at different temperatures. We take the small critical transition region formed by three or four neighboring $R_s(B)$ curves as one "critical" point. For instance, $R_c = 983.4$ Ω and $B_c = 0.9787$ T for a transition region from 20 mK to 30 mK. The scaling law can be further rewritten as $R_s(B,t)/R_c = F(|B - B_c|t)$, where parameter $t \equiv (T/T_0)^{-1/zv}$ and $T_0$ is the lowest temperature used in the scaling of the transition region. Ten representative "critical" points were selected and the results of normalized $R_s/R_c$ as a function of the scaling variable $|B - B_c|t$ are shown in Figs. S3-5. By adjusting the value of $t$, curves from various temperatures can collapse into one curve, and $zv$ can be extracted by linear fitting of $\ln(T/T_0)$ vs $\ln(t)$.

Moreover, the quantum Griffiths singularity (QGS) can be directly justified through the activated scaling law [3]. In finite temperature system, the influence of irrelevant parameter correction needs to be considered: $R = \Phi\left(\left(\frac{B-B_c^*}{B_c^*}\right) \cdot \left(ln\frac{T^*}{T}\right)^{\frac{1}{v\psi}}, u \cdot \left(ln\frac{T^*}{T}\right)^{-y}\right)$. Here $v$ and $\psi$ are critical exponents, $T^*$ is the characteristic temperature of quantum fluctuation, $B_c^*$ is the zero temperature critical field, $u$ is the leading irrelevant scaling variable and $y > 0$ is the associate irrelevant exponent. The activated scaling plots shown in Fig. 1(d), Fig. 2(d) and Fig. S7 are analyzed through the scaling procedure with irrelevant corrections as follows.

The irrelevant correction also influence the phase boundary $B_c(T)$ of superconductor-metal transition, with the crossing points satisfy [3]: $B_c(T) \propto B_c^*\left(1 - u \cdot \left(ln\frac{T^*}{T}\right)^{-\frac{1}{v\psi}-y}\right)$. The fitting curves of the phase boundary of superconductor-metal transition $B_c(T)$ are shown in the inset of Fig. 1(b), Fig. 2(b) and the inset of Fig. 3(c). The fitting parameters are summarized in Table S1. Considering the irrelevant scaling variable correction [4], the scaling function in the second argument [3] can be expanded as $R(B,T) = \Phi_1\left(\left(\frac{B-B_c^*}{B_c^*}\right) \cdot \left(ln\frac{T^*}{T}\right)^{\frac{1}{v\psi}}\right) + u \cdot \left(ln\frac{T^*}{T}\right)^{-y} \cdot \Phi_2\left(\left(\frac{B-B_c^*}{B_c^*}\right) \cdot \left(ln\frac{T^*}{T}\right)^{\frac{1}{v\psi}}\right)$, where both $\Phi_1$ and $\Phi_2$ are unknown arbitrary functions. For simplicity, we define $x_1(B,T) = (B - B_c^*) \cdot \left(ln\frac{T^*}{T}\right)^{\frac{1}{v\psi}}$ and $x_2(T) = \left(ln\frac{T^*}{T}\right)^{-y}$. Both $x_1$ and $x_2$ for definite $B$ and $T$ can be calculated using fitting parameters in Table S1. Thus, $R(B,T) = f(x_1) + x_2 \cdot g(x_1)$, where $f(x)$ and $g(x)$ are arbitrary functions. After a linear interpolation of $R(x_1)$ curves, we can obtain for any fixed $x_1 = x_{10}$ ($x_{10}$ is an arbitrary value of $x_1$), $R(T)|_{x_1=x_{10}} = f(x_{10}) + x_2(T) \cdot g(x_{10})$, which means $R(T)|_{x_1=x_{10}}$ and $x_2(T)$ are in linear relationship. The slope $k(x_{10})$ of $R(T)|_{x_1=x_{10}} - x_2(T)$ curve can be obtained from the linear fitting. We define corrected resistance $\tilde{R}(x_{10},T)$ for different temperatures as $\tilde{R}(x_{10},T) = R(T)|_{x_1=x_{10}} - k(x_{10}) \cdot x_2(T)$.

The perfectly coincident $\tilde{R}(x_1)$ curves shown in Fig. 1(d), Fig. 2(d) and Fig. S7 provide direct evidence of the quantum Griffiths phase.

In addition, the main difference between 4-ML and 6-ML PdTe$_2$ films is the strength of disorder, which can be revealed by the Ioffe-Regel parameter and mobility. The Ioffe-Regel parameter $k_F l$ for 4-ML and 6-ML films can be estimated by $k_F l = \frac{1}{ss'}\frac{2h}{e^2}\frac{1}{R_n}$ [5], where $s$ and $s'$ are spin degree of freedom and valley degree of freedom, h and e are Planck's constant and electron charge, and $R_n$ is the normal state resistance. Under the assumption of $s = s' = 2$, the Ioffe-Regel parameters for PdTe$_2$ films are summarized in Table S2, indicating that disorder strength in the 6-ML film is smaller than those of 4-ML films. Moreover, the mobility of 4-ML PdTe$_2$ film is around 0.119 $cm^2/(Vs)$ at 2 K, smaller than that of the 6-ML PdTe$_2$ film (around 0.865 $cm^2/(Vs)$). Therefore, compared to 6-ML film, the QGS is more clear and easier to be detected in 4-ML film with relatively strong disorder (large normal state resistance and small mobility).

Furthermore, we need to clarify that the disorder strength represented by Ioffe-Regel parameter only stands for the disorder of normal state fermions, while the disorder strength in the QGS represents disorder of local Cooper paring (which also renormalizes under coarse-graining when temperature is approaching zero). Even though, the Ioffe-Regel parameter may still provide a reference value for disorder strength in QGS.

The coherence length can be roughly estimated by two-dimensional (2D) Ginzburg-Landau (GL) formula, which can be expressed as follows [6]:

$$B_c^{\parallel} = \frac{\Phi_0 \sqrt{3}}{\pi \xi_{GL}(0) d} \sqrt{1 - \frac{T}{T_c}}$$

where $\Phi_0$ is the flux quantum, $\xi_{GL}(0)$ is the GL coherence length, and d is the thickness of the 2D superconductors (2.05 nm for 4-ML film and 3.08 nm for 6-ML film). According to the theoretical fittings based on 2D GL formula shown in Fig. S9(a), when the field is applied along the in-plane direction, the $\xi_{GL}(0)$ is estimated to be around 41.2 nm for 4-ML film and 50.2 nm for 6-ML film, much larger than the film thickness. Therefore, the vortex effect can be negligible under parallel magnetic field for both 4-ML and 6-ML PdTe$_2$ films. In the meanwhile, Fig. S9(b) presents that the temperature-dependent in-plane critical fields of ultrathin PdTe$_2$ films can be well fitted by the microscopic formula of Ising superconductivity [7], which also reveals the vortex effect can be negligible.

Under perpendicular magnetic field, the vortex-lattice phase can evolve into vortex-glass phase driven by quenched disorder, which finally leads to QGS. However, different from this scenario, under parallel field, the absence of vortex indicates a new microscopic mechanism for in-plane QGS. Under large in-plane magnetic field, the Zeeman splitting gives rise to pair breaking effect instead of vortex formation, which can also lead to a superconductor-metal transition. The quenched disorder effect influences the superconductor-metal transition, and results in the formation of large superconducting rare regions, which play a dominant role near the quantum critical point and finally lead to QGS. The above mechanism of QGS without vortex has been theoretically revealed in superconducting nanowires, which can be extended to 2D superconductors [8]. Moreover, since the ultrathin crystalline PdTe$_2$ films are type-II Ising superconductors [7], the pair-breaking effect under in-plane magnetic field is significantly affected by the effective Zeeman-type spin-orbit coupling as a function of disorder. To be specific, the regions with relatively large disorder are easier to lose superconductivity under in-plane magnetic field and form the normal state, while the other regions still keep superconducting and form rare regions. Therefore, a slightly inhomogeneous distribution of the disorder in PdTe$_2$ films can result in the formation of superconducting rare regions and finally give rise to the in-plane QGS.

## III. Figures and Tables

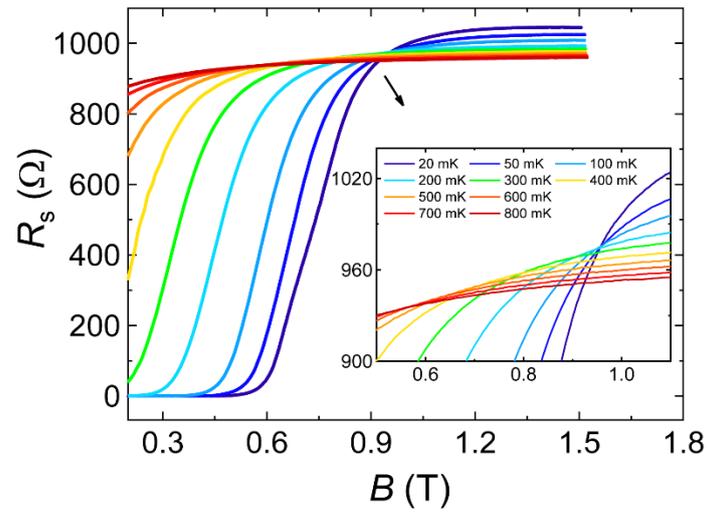

FIG. S1. The perpendicular magnetic field dependence of $R_s$ at various temperatures of the 4-ML PdTe$_2$ film. Inset: an enlarged view of the crossing region.

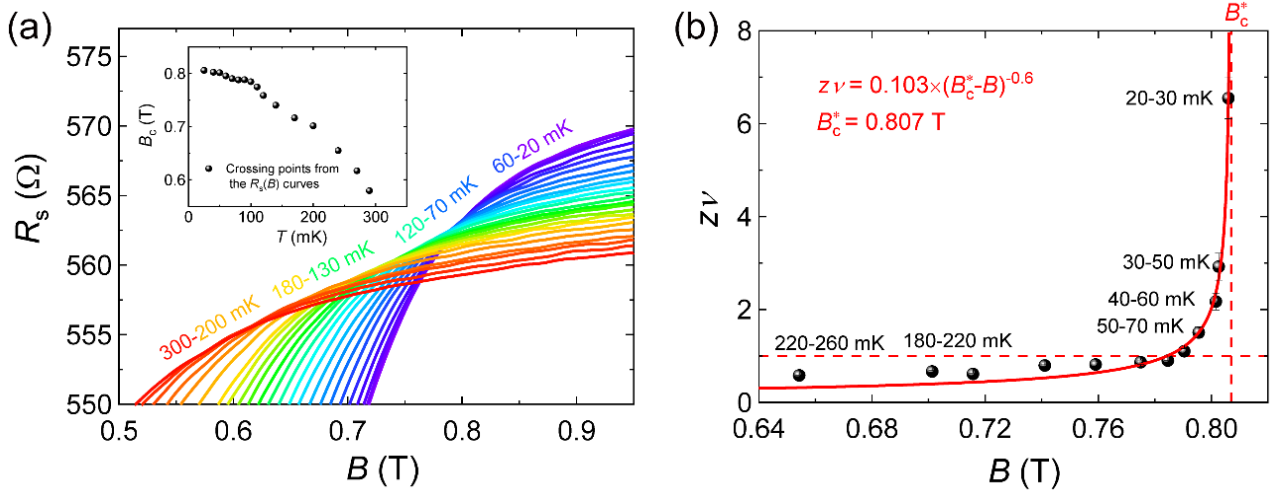

FIG. S2. The QGS of another 4-ML PdTe$_2$ film under perpendicular magnetic field. (a) The perpendicular magnetic field dependence of $R_s$ at different temperatures. Inset: Crossing points from the magnetoresistance isotherms. (b) Critical exponent $z\nu$ obtained from scaling analysis. The solid red line shows a fitting curve based on the activated scaling law and $B_c^* = 0.807$ T (vertical dashed line).

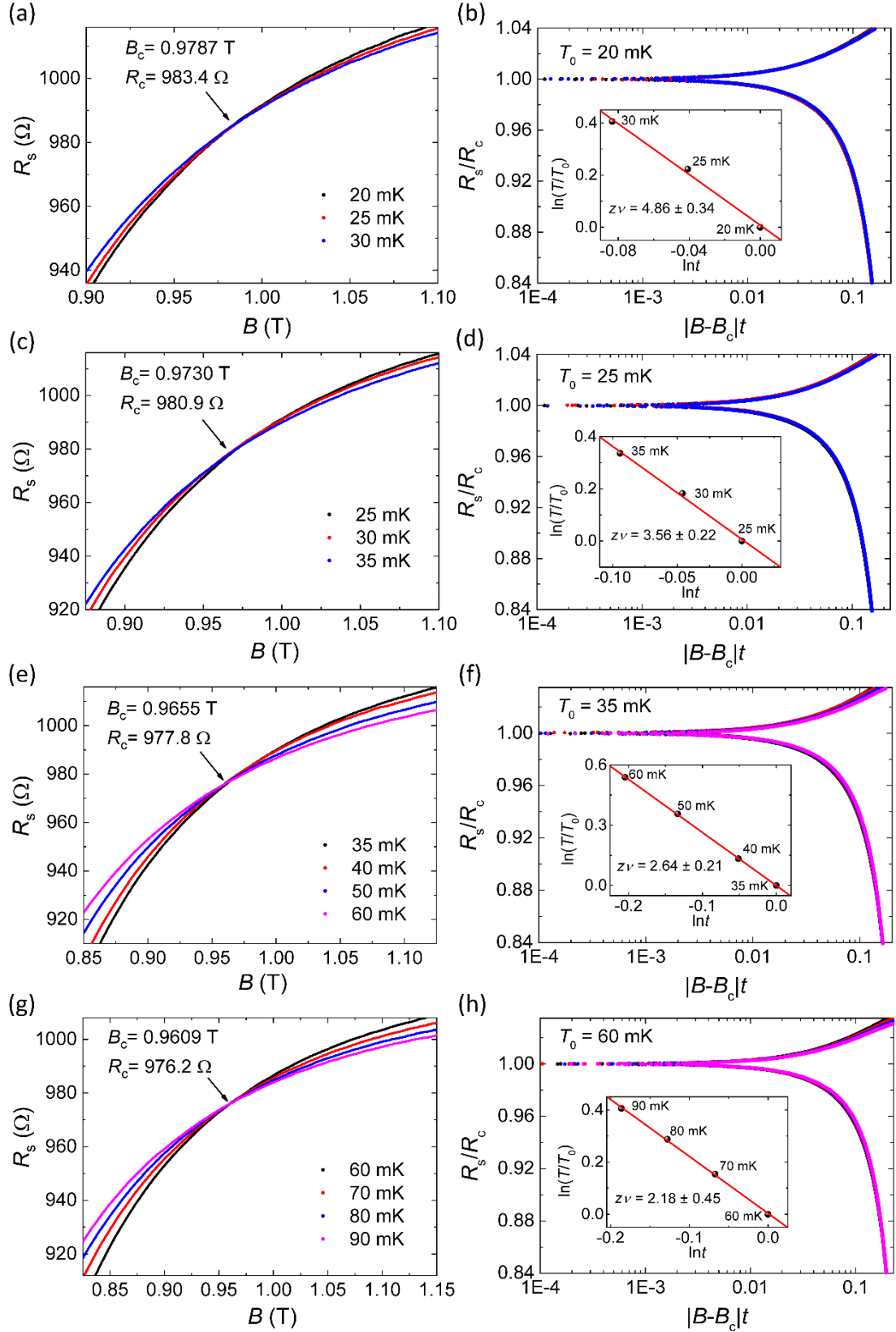

FIG. S3. Finite-size scaling analysis for the 4-ML PdTe$_2$ film at temperatures from 20 to 90 mK. (a) (c) (e) (g) Sheet resistance as a function of magnetic field at various temperature ranges of 20-30 mK (a), 25-35 mK (c), 35-60 mK (e) and 60-90 mK (g). (b) (d) (f) (h) Corresponding normalized sheet resistance as a function of scaling variable $|B - B_c|t$, with $t = T/T_0^{-1/zv}$. Inset: linear fitting between $\ln(T/T_0)$ and $\ln(t)$ gives effective "critical" exponent $zv$.

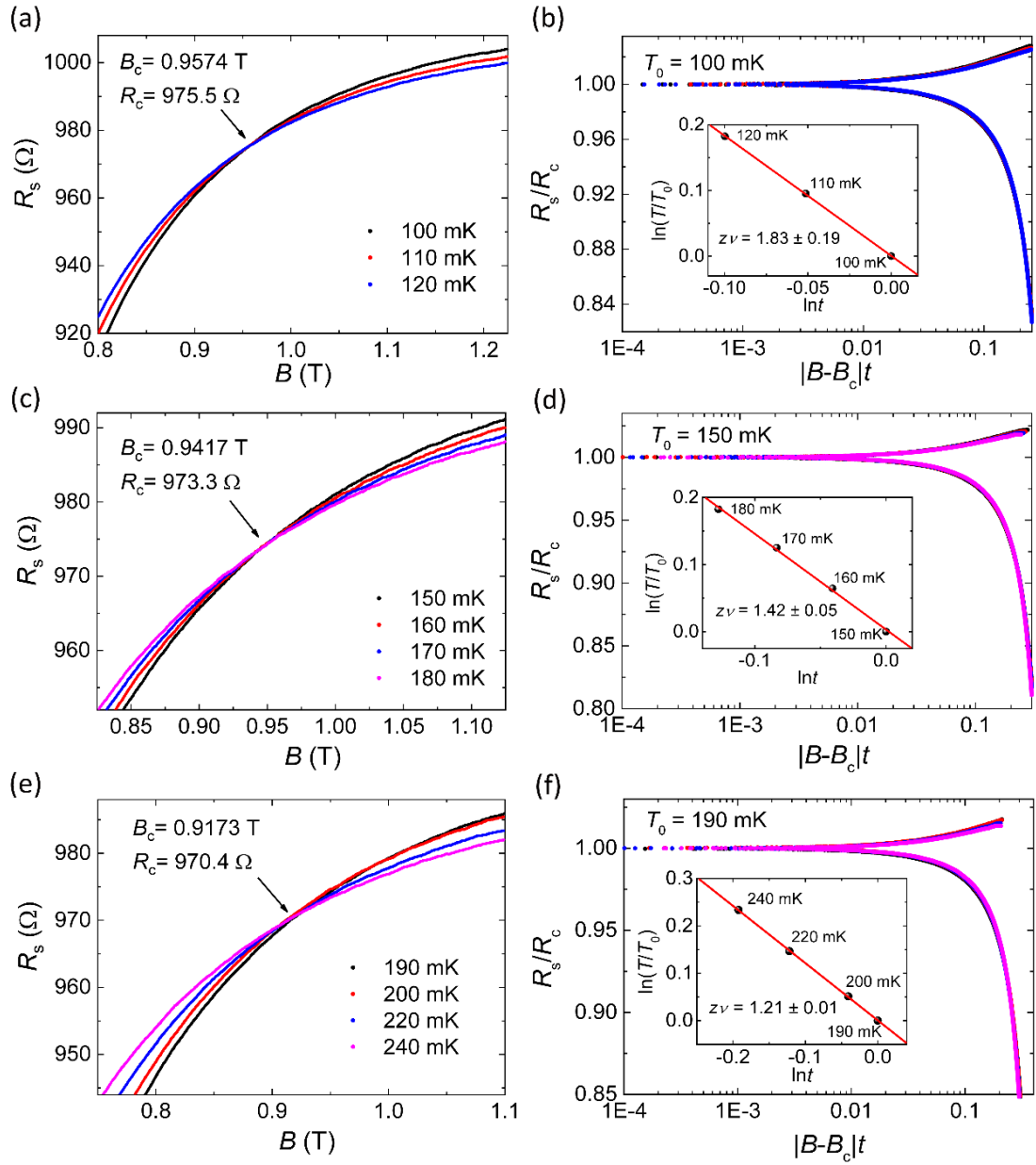

FIG. S4. Finite-size scaling analysis for 4-ML the PdTe$_2$ film at temperatures from 100 to 240 mK. (a) (c) (e) Sheet resistance as a function of magnetic field at various temperature ranges of 100-120 mK (a), 150-180 mK (c) and 190-240 mK (e). (b) (d) (f) Corresponding normalized sheet resistance as a function of scaling variable $|B - B_c|t$, with $t = T/T_0^{-1/zv}$. Inset: linear fitting between $\ln(T/T_0)$ and $\ln(t)$ gives effective "critical" exponent $zv$.

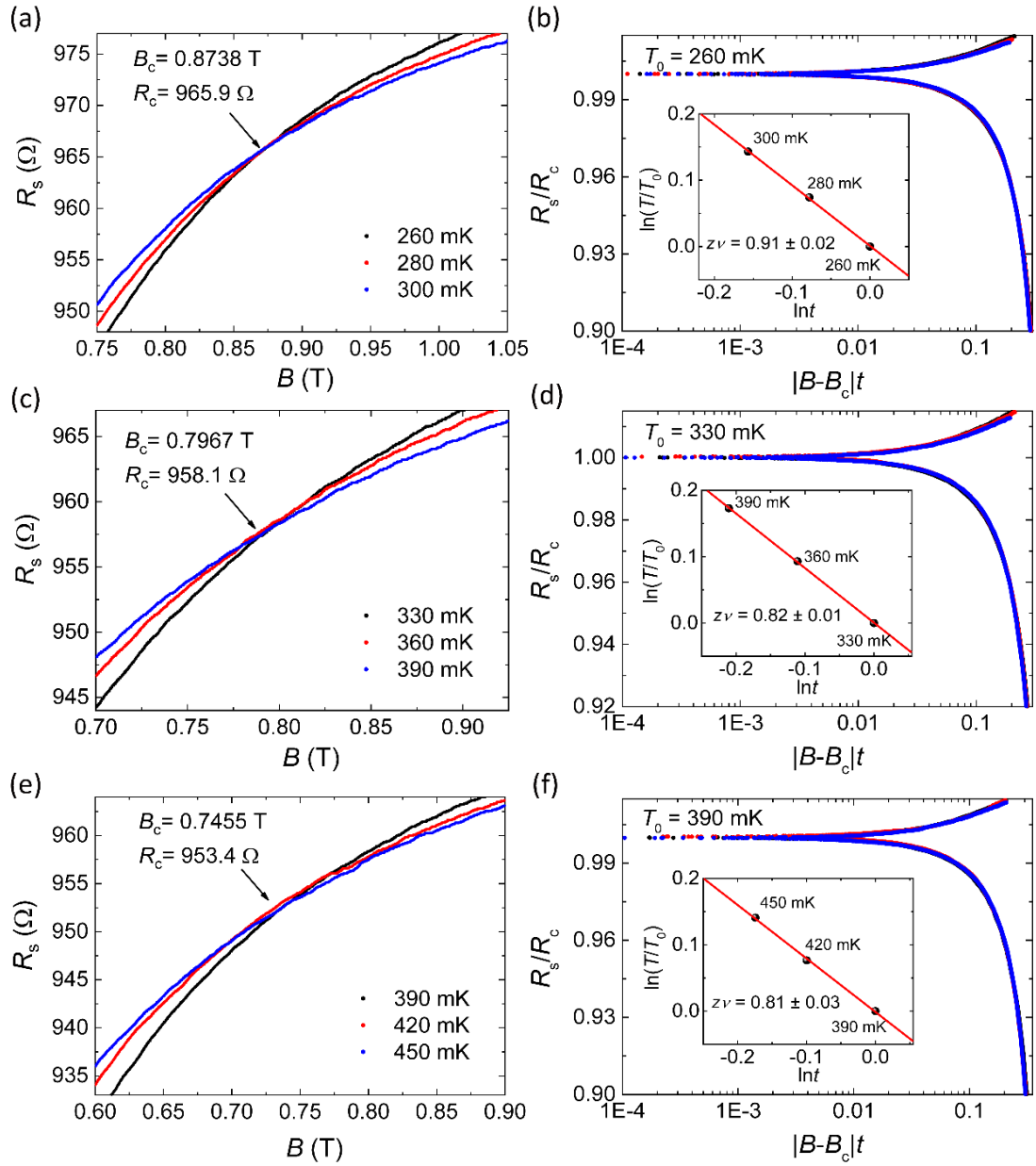

FIG. S5. Finite-size scaling analysis for the 4-ML PdTe$_2$ film at temperatures from 260 to 450 mK. (a) (c) (e) Sheet resistance as a function of magnetic field at various temperature ranges of 260-300 mK (a), 300-390 mK (c) and 390-450 mK (e). (b) (d) (f) Corresponding normalized sheet resistance as a function of scaling variable $|B - B_c|t$, with $t = T/T_0^{-1/zv}$. Inset: linear fitting between $\ln(T/T_0)$ and $\ln(t)$ gives effective "critical" exponent $zv$.

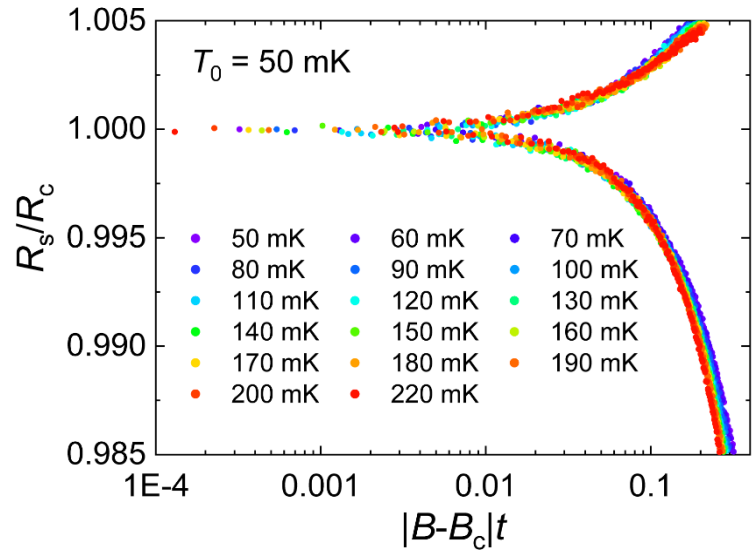

FIG. S6. Finite size scaling analysis for the 6-ML PdTe$_2$ film at temperatures from 50 to 220 mK under perpendicular magnetic field.

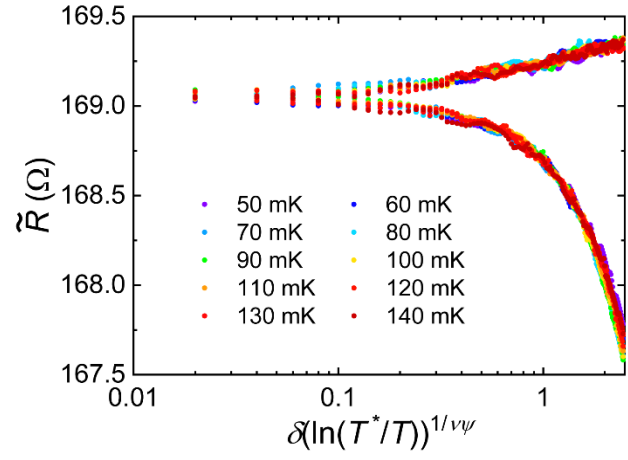

FIG. S7. The direct activated scaling analysis of 6-ML PdTe$_2$ film from 50 mK to 140 mK under parallel magnetic field. $\widetilde{R}$ represents the sheet resistance considering the irrelevant correction outlined in Part II.

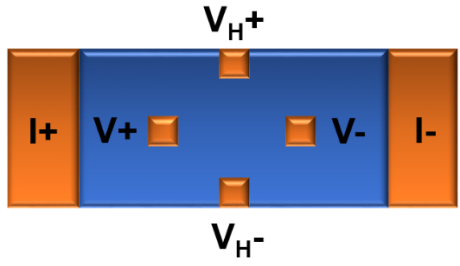

FIG. S8. The schematic for six-electrode transport measurement. $I+$ and $I-$ are current electrodes, $V+$ and $V-$ are voltage electrodes for longitudinal resistance, and $V_H+$ and $V_H-$ are Hall electrodes.

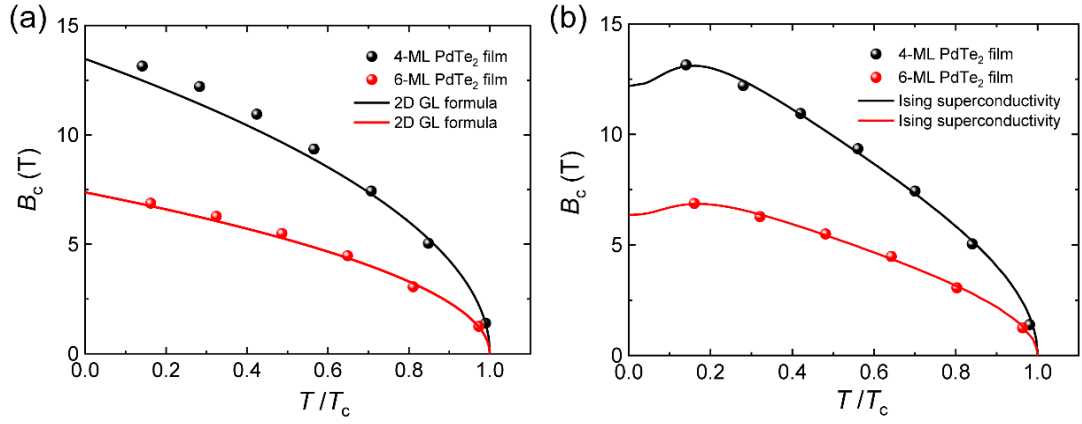

FIG. S9. The temperature dependence of the in-plane critical fields (black and red spheres, defined as the magnetic field corresponding to 50% of the normal state resistance) of PdTe$_2$ films and the theoretical fittings. (a) Fittings curves based on the 2D GL formula (black and red solid lines). The 2D GL formula describes the in-plane critical field near $T_c$, but may deviate from the experimental data in the low temperature regime. (b) Theoretical fittings based on the microscopic formula for Ising superconductivity (black and red solid lines).

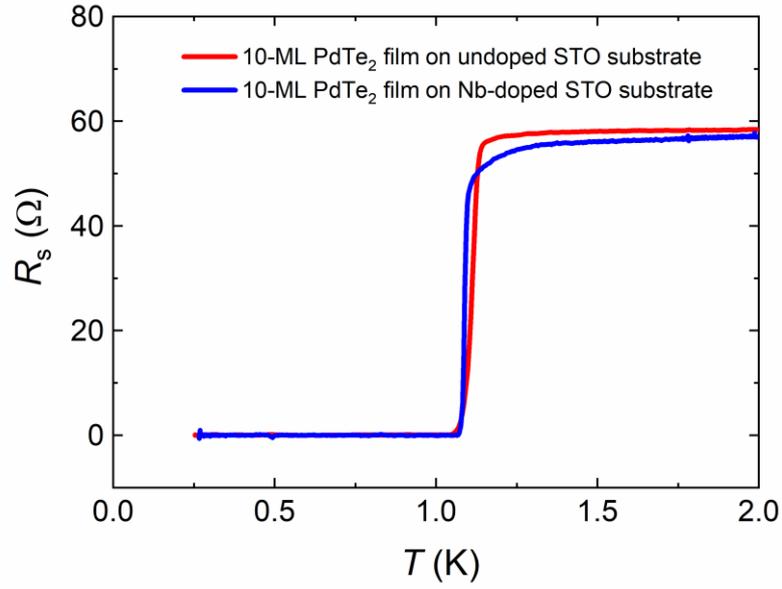

FIG. S10. The temperature dependence of sheet resistance at zero magnetic field for two 10-ML PdTe$_2$ films grown on undoped and Nb-doped STO substrates, respectively [7]. The normal state sheet resistance and superconducting transition temperatures are very similar for 10-ML PdTe$_2$ films on undoped and Nb-doped STO substrates, indicating that the Nb-doped STO does not contribute to the electronic transport results in our measurements. The excitation current flows through PdTe$_2$ films since there is a Schottky barrier between the film and the substrate.

Table S1. Fitting parameters of activated scaling analysis for PdTe$_2$ films under perpendicular and parallel fields.

|  | Thickness | $\nu\psi$ | $T^*$(mK) | $B_c^*$(T) | $y$ | $u$ |
|---|---|---|---|---|---|---|
| Perpendicular field | 4 ML | 0.6 | 1600 | 0.98 | 1.564 | 0.633 |
| Parallel field | 4 ML | 0.6 | 2000 | 15.28 | 1.615 | 0.308 |
| Parallel field | 6 ML | 0.6 | 2000 | 8.50 | 1.951 | 1.179 |

Table S2. Ioffe-Regel parameters for 4-ML and 6-ML PdTe$_2$ films.

| Sample thickness | 4 ML | 4 ML | 4 ML | 6 ML |
|---|---|---|---|---|
| Figure | Fig. 1 | Fig. 2 | Fig. S2 | Fig. 3 |
| Ioffe-Regel parameter | 14.5 | 9.62 | 25.6 | 82.0 |